\documentclass[a4paper,dvips,12pt]{article}
\usepackage{epsfig,graphics,graphicx}
\begin{document}
\pagestyle{myheadings} \markright{\it  CBPF-NF-006/03} \vskip.5in
\begin{center}
%
%
\vskip.4in {\Large\bf THE RELATIVISTIC DIRAC-MORSE PROBLEM VIA SUSY QM}
\vskip.3in
%
%
%
R. de Lima Rodrigues\footnote{Permanent address:
Departamento de Ci\^encias
Exatas e da Natureza, Universidade Federal de Campina Grande,
Cajazeiras - PB, 58.900-000-Brazil. \tt E-mail to RLR is
rafaelr@cbpf.br or rafael@fisica.ufpb.br,
the \tt e-mail to ANV is vaidya@if.ufrj.br.}\\
Centro Brasileiro de Pesquisas F\'\i sicas (CBPF)\\
Rua Dr. Xavier Sigaud, 150, CEP 22290-180, Rio de Janeiro-RJ, Brazil\\
A. N. Vaidya\\
Instituto de F\'\i sica - Universidade Federal do Rio de Janeiro\\
Caixa Postal 68528 - CEP 21945-970, Rio de Janeiro, Brazil
%
%
%
\end{center}

\vskip.2in
\begin{abstract}
The Morse problem
is investigated in relativistic quantum mechanics.
\end{abstract}

\vspace{1cm}

\vspace{1cm}

Talk at the Brasilian National Meeting on Particles and Fields, October 15
to 19, 2002, \'Aguas de Lind\'oia-SP, Brazil. To appear in the proceedings,
site www.sbf1.if.usp.br/eventos/enfpc/xxiii. Preprint CBPF-NF-006/03, site
cbpf.br.

\section{Introduction}

The Dirac oscillator was first formulated and investigated in the
context of nonrelativistic quantum mechanics
by Moshinsky-Szczepaniak \cite{Mosh89}. They construct
a Dirac Hamiltonian, linear in the momentum $\vec p$ and position $\vec r,$
whose square leads to the ordinary harmonic oscillator in the
nonrelativistic limit.
In 1992, Dixit {\it et al.} instead of modifying the momentum,
considered a
Dirac oscillator with scalar coupling, by modification of the mass term
\cite{Dixit92}.

In a recent paper Alhaidari has proposed a generalization for Moshinsky's
equation which contains the
oscillator and the Coulomb
problem as special cases.
He also
claims to have formulated and solved the
Dirac-Morse problem and obtaining its relativistic bound states
and spinor wave functions \cite{Alhai}.
In a later paper a similar method is applied to include shape invariant
potentials via supersymmetry in quantum mechanics (SUSYQM) \cite{inv01}.
However, his calculation  contains serious mistakes as we
explain below \cite{VR02a}.
We also explain another way of treating the problem.

\section{Problem with Alhaidari's Calculation}

The Hamiltonian that appears in the work of Alhaidari \cite{Alhai}
does not lead to his
equation (1). The Hamiltonian should be
in the notation is that of Bjorken and Drell \cite{BD}.

\begin{equation}
H= {\mbox{\boldmath $\alpha$}}\cdot({\bf p}-i\beta{\bf {\hat  r}}W(r)) +
\beta M+V(r)
\end{equation}
where ${\bf {\hat r}} = {{\bf r}\over r}$ and we consider
$\hbar=1$ and the mass as being $M.$
 Due to the matrix $\beta$ accompanying $W$ in the Hamiltonian Alhaidari's
interpretation of the vector $ (V,{\bf {\hat r}} W)$ as an
electromagnetic potential is incorrect.
To verify our assertion we put

\begin{equation}
\Psi=\pmatrix{{iG_{\ell j}\over r}\phi^{\ell}_{jm}\cr {F_{\ell j}\over r}
{{\mbox{\boldmath $\sigma$}}\cdot{\bf {\hat r}}}\phi^{\ell}_{jm}\cr}
\end{equation}
where $ \phi^{\ell}_{jm} $ are as defined in reference \cite{BD}.
Then using the relations

\begin{equation}
{\mbox{\boldmath $\sigma$}}\cdot{\bf p}{f(r)\over r}\phi^{\ell}_{jm}=-
{i\over r}({df\over dr}+{\kappa f\over r}){\mbox{\boldmath $\sigma$}}
\cdot{\bf{\hat r}}\phi^{\ell}_{jm}
\end{equation}
and

\begin{equation}
{\mbox{\boldmath $\sigma$}}\cdot{\bf p}{\mbox{\boldmath $\sigma$}}
\cdot{\bf{\hat r}}{f(r)\over r}\phi^{\ell}_{jm}=-{i\over r}({df\over dr}-
{\kappa f\over r})\phi^{\ell}_{jm}
\end{equation}
where $\kappa=\pm (j+{1\over 2})$ for $\ell=j\pm{1\over 2}$,
and defining

\begin{equation}
\Phi=\pmatrix{G_{\ell j}\cr F_{\ell j}\cr}
\end{equation}
we get the radial equations
\begin{equation}
\left(-i\rho_2{d\over dr} +\rho_1(W+{\kappa\over r})-(E-V)+M\rho_3\right)\Phi=0
\end{equation}
where $\rho_i$ are the Pauli matrices. The last equation corresponds
to Alhaidari's equation (1) where the quantum numbers $l$ and $j$ are
omitted incorrectly.

 Next, there is no reason for the functions $V(r)$ and $W(r)$
which appear in the Hamiltonian to depend on the angular quantum numbers
which make their appearance only when we separate variables to solve the
Dirac equation. Hence his choice of the constraint

\begin{equation}
W(r)={\alpha\over S}V(r)-{\kappa\over r}
\end{equation}
cannot be satisfied since $\kappa=\pm (j+{1\over 2})$ for $l=j\pm{1\over 2}$
is not a fixed quantity.
Since the Dirac operator

\begin{equation}
K={\gamma}^0(1+{{\mbox{\boldmath $\Sigma$}}\cdot{\bf L}})
\end{equation}
satisfies the property

\begin{equation}
K\phi^{\ell}_{jm}=-\kappa \phi^{\ell}_{jm}.
\end{equation}
Alhaidari could have avoided the contradiction by taking the Hamiltonian to be

\begin{equation}
H= {\mbox{\boldmath $\alpha$}}\cdot({\bf p}-i\beta{\bf {\hat r}}(W(r)+
{K\over r})) +\beta M+V(r)
\end{equation}
which leads to the radial equations

\begin{equation}
\left(-i\rho_2{d\over dr} +W\rho_1-(E-V)+M\rho_3\right)\Phi=0.
\end{equation}
Applying the transformation

\begin{equation}
\Phi=e^{-{i\rho_2 \eta}}\hat\Phi
\end{equation}
we have

\begin{eqnarray}
{}&&\left[-i\rho_2{d\over dr}-(E-V)+\rho_1(W\cos 2\eta-M\sin 2\eta)\right]\hat\Phi
\nonumber\\
{}&&+
\left[\rho_3 (W\sin 2\eta+M\cos 2\eta)\right]\hat\Phi=0.
\end{eqnarray}
Choosing
\begin{eqnarray}
W={V\over \sin 2\eta}.
\end{eqnarray}
Hence

\begin{eqnarray}
{}&&\left[-i\rho_2{d\over dr}-E+V(1+\rho_3)+\rho_1({V\over \tan 2\eta}-
M\sin 2\eta)\right]\hat\Phi\nonumber\\
{}&&+
\left[\rho_3 M\cos 2\eta\right]\hat\Phi=0
\end{eqnarray}
which gives

\begin{equation}
\label{FG}
\hat F_{\ell j}={1\over {E+M\cos 2\eta}}({d\over dr}+{V\over \tan 2\eta}-
M\sin 2\eta)\hat G_{\ell j}
\end{equation}
and

\begin{equation}
\left[-{d^2\over dr^2}+({V\over \tan 2\eta})^2+2EV-
{1\over \tan 2\eta}{dV\over dr}-
(E^2-M^2)\right]\hat G_{\ell j}=0.
\end{equation}
The last three equations correspond to equations (3-5) of Alhaidari.
The choice $W={V\over \sin 2\eta}$ gives equations (4) and (7) of
Alhaidari.

The resulting energy levels will be degenerate in $l, j, m$
in contrast to what happens in the case of relativistic Coulomb and
relativistic harmonic oscillator problems.

\section{The Morse potential: An alternative treatment}

Even if  one interprets the results of Alhaidari as
corresponding only to $\ell=0, s=0, k=-1,$ the unitary transformation
is puzzling. It does not appear at all if we start with $V=0$ as in the
treatment of Casta\~nos {\it et al.} earlier  \cite{Cas91}. Then equation
(\ref{FG}) gives $F$ in terms of $G$ and the second order equation for
 $k=-1$ is

\begin{equation}
\left[-{d^2\over dr^2}+(W-{1\over r})^2+\frac{d}{dr}(W-{1\over r})
-(E^2-M^2)\right]\hat G_{\ell j}=0.
\end{equation}

The  choice  $W={1\over r}+A-Be^{-\lambda r}$
gives the S-wave Morse potential in quantum mechanics
with an additional $A^2$ term in the potential.
Although the relativistic spectrum is different,
the same non-relativistic limit is obtained.

\section{Conclusion}

In this communication we have pointed out the limitation
of a published treatment of the relativistic Morse potential
problem. We also present an alternative treatment which is similar to
the supersymmetric formulation of the Dirac oscillator as given
by Cast\~nos {\it et al.} \cite{Cas91}.
A similar modification holds for relativistic shape invariant
potentials as well.

\vspace{1cm}
\centerline{\bf ACKNOWLEDGMENTS}

RLR was supported in part by CNPq (Brazilian Research Agency). He wishes to
thank J. A. Helayel Neto for the kind of hospitality at CBPF-MCT. The
authors wish also thank the staff of the CBPF and DCEN-CFP-UFCG. The
authors are also grateful to the organizing commitee at the
 Brazilian National Meeting on Particles and fields,
October 15 to 19, 2002, \'Aguas de Lind\'oia-SP, Brazil.

\end{document}